# Considering pricing and uncertainty in designing a reverse logistics network

Mohsen Zamani[1], Mahdi Abolghasemi[2], Seyed Mohammad Seyed Hosseini[3], Mir Saman Pishvaee[4]

[1]Mohsen Zamani, Department of Industrial engineering, School of Industrial Engineering, Iran University of Science and Technology, Ave. Narmak, 16846-13114- Tehran, Iran,
Email: mohsen_zamani@alumni.iust.ac.ir
*[2] Corresponding Author: Mahdi Abolghasemi, School of Mathematical and Physical Sciences, University of Newcastle, Newcastle, NSW, Australia, 2308, Email: mahdi.abolghasemi@uon.edu.au
[3]Seyed-Mohammad Seyed-Hosseini, Department of Industrial engineering, School of Industrial Engineering, Iran University of Science and Technology, Ave. Narmak, 16846-13114- Tehran, Iran,
Fax: (+98)21 73225007, Email: seyedhosseini@ iust.ac.ir
[4]Mir Saman Pishvaee, Department of systems and supply chain engineering, School of Industrial Engineering, Iran University of Science and Technology, Ave. Narmak, 16846-13114- Tehran, Iran
Fax: (+98)21 73225016, Email: pishvaee@iust.ac.ir

**Biographical notes:**
Mohsen Zamani holds MSc in Industrial engineering from Iran University of Science and Technology (IUST), Iran. His research interest is considering risk in closed loop supply chain management. He is working as supply chain experts in the car manufacturing industry.

Mahdi Abolghasemi is a PhD candidate in Statistics and casual academic at the university of Newcastle, Australia. He holds a bachelor and master in Industrial engineering from Bu-Ali Sina university, Iran and a MSc in Industrial Management from Madrid Poly Technique University, Spain. His main researches interests are statistical modelling, applied machine learning, operation management, and predictive analytics.

Seyed Mohammad Seyed Hosseini is a Professor at Department of Industrial Engineering, School of Industrial Engineering, Iran University of Science and Technology. The Editor in chief of international journal of industrial engineering and production research, earned his Doctor of Philosophy in Industrial and Civil Engineering (Transportation Planning) from the University of Oklahoma, Norman, in 1982 and is the author of numerous works in the field of Production Management, Operation Management, Location Theory, Plant Design and Quality Engineering.

Mir Saman Pisvaee is an Associate Professor at Department of systems and supply chain engineering, School of Industrial Engineering, Iran University of Science and Technology. He has been awarded Doctor of Philosophy (Ph.D.) in Industrial Engineering, University of Tehran, 2012. He has been the author of several publications in the field of Supply chain management,

Robust optimization, Sustainable operations management and International logistics Uncertain programming.

**Abstract**

Companies try to maximize their profits by recovering returned products of highly uncertain quality and quantity. In this paper, a reverse logistics network for an Original Equipment Manufacturer (OEM) is presented. Returned products are selected for remanufacturing or scrapping, based on their quality and proportional prices are offered to customers. A Mixed Integer Non-linear Programming (MINLP) model is proposed to determine the location of collection centers, the optimum price of returned products and the sorting policy. The risk in the objective function is measured using the Conditional Value at Risk (CVaR) metric. CVaR measures the risk of an investment in a conservative way by considering the maximum lost. The results are analyzed for various values of the risk parameters ($\alpha$, and $\lambda$). These parameters indicate that considering risk affects prices, the classification of returned products, the location of collection centers and, consequently, the objective function. The model performs more conservatively when the weight of the CVaR part ($\lambda$) and the value of the confidence level $\alpha$ are increased. The results show that better profits are obtained when we take CVaR into account.

**Keywords**: Reverse logistics, Network design, Pricing, Risk.

## 1. Introduction

Reverse logistics is "*the process of planning, implementing, and controlling the efficient, effective inbound flow and storage of secondary goods and related information opposite to the traditional supply chain direction for the purpose of recovering value and proper disposal*''(Fleischmann, 2001). In recent years, for legal, economic and environmental reasons, governments and companies have paid more attention to reverse supply chains. Clients and stakeholders impose a great deal of pressure on companies to design eco-friendly products (Nenes and Nikolaidis, 2012). Deterioration of the environment, conservation of energy resources, and waste disposal issues have led companies to recover or refurbish returned products instead of merely mass producing new items (Takahashi et al., 2012). Legal and economic incentives are the driving factors for implementing Closed-Loop Supply Chains (CLSC) (Minner and Kiesmüller, 2012). CLSC is defined by Guide Jr and Van Wassenhove (2009) to be the "*design, control, and operation of a system to maximize value creation over the entire life cycle of a product with dynamic recovery of value from different types and volumes of returns over time,*" which helps industries to remain sustainable. This will become more important in the near future due to growing concerns over the depletion of natural resources (Carrasco-Gallego et al., 2012).

In some cases, remanufacturing returned products can reduce manufacturing costs by about 40 to 60 percent. Utilizing used components from returned products can reduce manufacturing costs by up to 80 percent (Dowlatshahi, 2000). Furthermore, combining remanufacturing with production can increase market share while sustaining profit margins (Ferrer, 1997).

Collecting returned products is the first step in recovery operations and plays a vital role in all recovery processes. The quality of returned products, the prices offered, and the location of collection centers are pivotal factors in designing reverse supply chains. Some companies collect their returned products themselves, while some others delegate this task to third-party companies (Kaya, 2010). Offering reasonable prices for returned products also plays a key role in enabling companies to collect appropriate volumes of products to meet their recovery aims (Daniel et al., 2000).

The most common factors encouraging customers to take returned products back to the OEM, rather than local remanufacturing companies are the prices offered and the accessibility of collection centers. In some situations, trade-in rebates (when customers return existing products at the same time as purchasing new products) offer a financial stimulus to product holders, acting as an incentive for them to return their used goods (Aras and Aksen, 2008). In addition, companies should establish collection centers as close as possible to demand nodes in order to facilitate the collection of returned products. The recovery of returned products is more cost-effective if the quantity of returned goods is sufficient, and the quality of the goods is almost the same as that required by potential customers. However, the quality of returned products is highly variable and has a direct impact on recovery activities, particularly on the prices offered (Galbreth and Blackburn, 2006).

Uncertainty is an unavoidable challenge in CLSC. The price, quality and quantity of returned products are some types of uncertainty that could occur in a reverse supply chain. In this paper, we use the CVaR metric to evaluate the risks for companies employing reverse supply chains. The main contribution of this paper are determining the optimal locations of collection centers, sorting returned products according to their quality and determining the optimal incentive price to be offered for returned goods after taking risk into account. The remainder of this paper is organized as follows. Section 2 will review the literature. In section 3 the model is defined. A numerical example is presented in section 4 to demonstrate the validity of the model presented. An analysis of the model's performance and our conclusions are presented in Sections 5 and 6, respectively.

## 2. Literature review

Various authors have investigated the development and solution of CLSC problems. Some researchers have concentrated their attention on decision variables such as the price and location of collection centers. Other factors that arise in the consideration of reverse logistic activities are the collection, recovery and disposal of returned products. Previous models have incorporated the demand for and uncertainty of the quality of returned goods. Two solution techniques that are commonly employed are game theory and the consideration of cost functions. In this section, we review some of the papers appearing in the literature in which these factors have been discussed.

### 2.1. Network design

Min et al. (2006) presented a MINLP model and genetic algorithm (GA) solution for designing a reverse logistics network. Their model considered inventory and shipment costs, and freight discounts. They also analyzed interplays between the initial collection point and centralized return centers.

Alshamsi and Diabat (2015) stated that reverse logistics is initiated at the consumer level with the collection of used items. They presented a mixed integer linear programing model which addressed the network design problem and determined the best locations for collection centers, and the optimal capacities of inspection and remanufacturing facilities. Their main contribution was the proposal of "in-house fleeting", in which an onsite fleet manager is employed to deal with issues more efficiently, as an outsourcing option for the collection of used products. They asserted that the initial investment on fleet or center expansion must be taken into account, and the liable parties should determine the process for the return of products and the location of return centers. Their model is effective for companies that are responsible for the collection of huge volumes of products and deal with large-scale transportation systems.

Diabat et al. (2013) surveyed reverse logistic networks and determined the optimal numbers and locations of initial and central collection centers as well as the maximum holding times for returned products. Rachaniotis et al. (2010) noted that the environmental aspects of reverse logistics might force companies to reuse returned products rather than waste them. They took budget constraints, the environmental savings from re-assembling used products and multi period facilities layouts into account, and obtained Pareto optimal solutions for the overall performance of these networks.

Kannan et al. (2012) investigated the impact of government regulations related to collecting End-Of-Life (EOL) products. They showed that these regulations have forced companies to focus on acquiring used items. Their model incorporated environmental emissions into the cost function.

Diabat et al. (2015) considered Distribution Centers (DCs) and Remanufacturing Centers (RCs). In their model, RCs are responsible for returning remanufactured products to the retailers that are in direct contact with the DCs through the supply chain. Their computational results showed a reduction in the number of established facilities in the supply chain as the inventory cost increases. On the other hand, they found that an increase in transportation costs led to a rise in the number of established return facilities. The influence of the acquisition price paid to the customer was not considered in their model.

### 2.2. Incorporation of the pricing in network design

A combination of pricing and network design are the main issues considered in a number of studies. Aras and Aksen (2008) considered the location of collection centers and the financial incentives offered to customers. They used a Drop-off strategy in which customers travel from demand nodes to collection centers in order to return their products. The quality of returned products was classified into different levels and differing financial incentives were offered for returns of each of the quality types.

Kaya (2010) studied a system which included a manufacturer that both produced new products and remanufactured returned products. He also considered three different scenarios, including both centralized and decentralized settings for the return of products. He found an optimal policy for pricing returned products and also optimal production levels for new products. He showed that price and distance to collection centers were the most important incentives for customers to return goods and presence of local remanufacturers can have an important influence on customer behavior. That is, local remanufacturers could absorb returned products by providing privilege to customers.

In some cases, local remanufacturers are also considered to be cooperators or competitors. Jung and Hwang (2011) announced that a reverse logistic system consists of an OEM and a remanufacturer. New products are sold under the "take-back requirement" policy which means that a penalty is imposed on OEMs if a take-back quota for the end-of-use products is not met. They determined a pricing policy for the OEM and the remanufacturer under policies of competition and cooperation.

Sorting returned products plays a crucial role in obtaining the best pricing policy. A number of studies optimized pricing with regard to the quality of the returned products. Nenes and Nikolaidis (2012) surveyed sorting policies. They considered multi-period policies, pricing based on several quality levels, and differing remanufacturing and stocking operations for companies. They originally presented a model for remanufacturing cell phones, but asserted that it could be generalized to other systems. Indeed, El Saadany and Jaber (2010) studied Economic Production Quantity (EPQ) models and found the best strategy for pricing returned products with respect to their quality levels. They presented numerical examples which demonstrated that either total disposal or total remanufacturing is the best policy.

### 2.3. Consideration of the uncertainty

It has been suggested that, for non-repetitive decision making problems incorporating uncertainty, a risk-averse approach that considers the effects of the variability of random outcomes, such as the random total cost, will provide more robust solutions than a risk-neutral approach (Noyan, 2012). Pokharel and Liang (2012) proposed a model for determining the quantity and acquisition price of returned products. Managerial aspects were not considered in their model. Financial incentives affected the number of returned products, and the remanufacturing cost depended on the quality of the returned products, which was uncertain. They assumed that the incentive price was determined on the basis of historical or forecasted returns for each level of quality of the returned products.

Alhaj and Diabat (2010) added uncertainty into their model by incorporating a new variable which reflected the probability of different scenarios occurring. They considered different demand scenarios, and compared the results in deterministic and stochastic form by solving deterministic models for each scenario. They concluded that uncertainty had a significant impact on the total cost to the OEM.

Diabat et al. (2009) proposed a Capacitated warehouse Location Model with Risk Pooling (CLMRP). Their model consisted of a supplier and multiple retailers which dealt with stochastic demand. They allowed some retailers to act as distribution centers. Their main aim was to present a GA to solve a CLMRP rather than to assess the economic parameters associated with their model.

Many metrics, including CVaR, have been employed in financial optimization over recent years (Markowitz (1952), Krokhmal et al. (2002), Benati (2003), and Yau et al. (2011)). However, incorporating CVaR into reverse logistics is a fairly recent development. This measure was initially employed to solve financial problems (Rockafellar and Uryasev, 2000).

Soleimani and Govindan (2014) presented a risk-averse, two-stage stochastic model for designing a reverse supply chain. They used CVaR in their model, which considered the prices and quantities of returned products. By linearization of CVaR, they determined the optimum point in the two-stage stochastic structure using a mean risk objective function. An appropriate example was then designed and analyzed.

An approach similar to that of Soleimani and Govindan (2014) was presented for CLSC by Soleimani et al. (2014). They incorporated the CVaR, Value at Risk (VaR), and Mean Absolute Division (MAD) metrics into a two-stage stochastic programming model. As they were studying CLSC, the drawbacks of a reverse loop could be determined from the profitability of a forward supply chain. In addition, a pure reverse logistics network was designed, developed, and formulated. Moreover, extensive evaluations of the capabilities of a mean-risk model for the proposed network were undertaken. Jena and Sarmah (2016) considered a model with common retailer and uncertain demand in which price and service level like warranty and maintenance were the most important factors to determine customer's tendency. They considered four different configurations including global system where both the manufacturer and retailer act together, integrated system where each manufacturer cooperates separately with the retailer, decentralized system where each member tries to maximize its own profit regardless of the others and direct system where products are sold directly to the market and end user customers can purchase them directly from manufacturers. They found that when demand variation increases, direct system is more profitable than the other channels. They also showed that total channel profit in global system is more than other channels and a direct relationship between market size and total profit in all channels. They did not take location and transportation cost into account. Keyvanshokooh et al. (2013) designed and resolved a forward/reverse logistics network where return products were categorized with respect to their quality level and different acquisition price. They considered a uniform distribution for the expected price of one unit of used products by customers and found the acquisition price using a dynamic pricing approach. Their main contribution was using dynamic pricing to find acquisition price and using that in a MILP model for further decisions in network. Table 1 summarizes the findings from the literature on different approaches to pricing and network design in reverse logistics as well as the model presented in this paper.

Table 1. Some of the related papers in pricing and network design of reverse logistics.

| Reference articles | Network design | Competition | Uncertainty | | Reverse logistics activity | | | Design variable | | Approach of dealing with | |
|---|---|---|---|---|---|---|---|---|---|---|---|
| | | | Demand | Quality | Disposal | Recovery | Collection | Location | price | Game theory | Cost function |
| (Keyvanshokooh et al., 2013) | ✓ | | ✓ | ✓ | ✓ | ✓ | ✓ | ✓ | ✓ | | ✓ |
| (Wu, 2013) | | ✓ | | | | ✓ | | | ✓ | ✓ | |
| (Qiang et al., 2013) | ✓ | ✓ | ✓ | | | | ✓ | | ✓ | ✓ | |
| (Wu, 2012) | | ✓ | | | | ✓ | | | ✓ | | ✓ |
| (Pokharel and Liang, 2012) | ✓ | | ✓ | ✓ | | ✓ | ✓ | | ✓ | | ✓ |
| (Nenes and Nikolaidis, 2012) | ✓ | | | ✓ | ✓ | ✓ | | | | | ✓ |
| (Minner and Kiesmüller, 2012) | | | | | | ✓ | | | ✓ | | ✓ |
| (Jung and Hwang, 2011) | | ✓ | | | | ✓ | | | ✓ | | ✓ |
| (Shi et al., 2011) | | | | | | ✓ | | | ✓ | | ✓ |
| (Teunter and Flapper, 2011) | | | ✓ | ✓ | | ✓ | ✓ | | ✓ | | |
| (Ferrer and Swaminathan, 2010) | | | | | | ✓ | | | ✓ | | ✓ |
| (Kaya, 2010) | | | ✓ | | | ✓ | ✓ | | ✓ | | ✓ |
| (El Saadany and Jaber, 2010) | | | | | | ✓ | ✓ | | ✓ | | ✓ |
| (Pellerin et al., 2009) | | | | | | ✓ | | | | | ✓ |
| (Aras and Aksen, 2008) | ✓ | | | ✓ | | ✓ | ✓ | ✓ | ✓ | | ✓ |
| (Tagaras and Zikopoulos, 2008) | ✓ | | ✓ | ✓ | | ✓ | ✓ | | ✓ | | ✓ |
| (Atasu et al., 2008) | | ✓ | | | | ✓ | | | ✓ | ✓ | |
| (Galbreth and Blackburn, 2006) | | | ✓ | ✓ | ✓ | ✓ | | | | | ✓ |
| (Bakal and Akcali, 2006) | | | | | | ✓ | | | ✓ | | ✓ |
| (Ferguson and Toktay, 2006) | | ✓ | | | | ✓ | | | ✓ | ✓ | |
| (Robotis et al., 2005) | | | | ✓ | | ✓ | | | | | |
| (Guide et al., 2003) | | | ✓ | | | ✓ | | | ✓ | | ✓ |
| (Klausner and Hendrickson, 2000) | | | | | | ✓ | | | ✓ | | ✓ |
| This paper | ✓ | ✓ | ✓ | ✓ | | ✓ | ✓ | ✓ | ✓ | | ✓ |

As seen in Table 1, a small section of the literature associated with reverse logistics investigated the integrated network design and pricing problem. It must be noted that the majority of the presented models did not consider the cooperation of customers in returning products.

## 3. Model description

Remanufacturing is the most profitable recovery activity that can be performed for returned products. Collecting high-quality returned products is vital in a competitive market, since remanufacturing them can increase the profits of a company. On the other hand, remanufacturing low quality products is not economical. It would be better to sell them as scrap materials. The purpose of this paper is to provide a model for the collection of returned products that considers quality for an OEM in a competitive market. Returned products are categorized into two types, based on their quality, and designated either for remanufacturing or for scrapping. Theoretically, it is possible that all returned products can be remanufactured except for the low quality ones for which remanufacturing costs are prohibitively high, making them more suitable for selling as scrap.

Prices, based on the quality level, are offered to customers in various returning locations. The quantity of returned products collected is based on the utility (facilities offered) of local remanufacturers. It is presumed that returned products are scrapped by selling them in collection centers, while remanufacturing requires the shipment of returned products to recovery centers. The network of facilities is presented in figure 1.

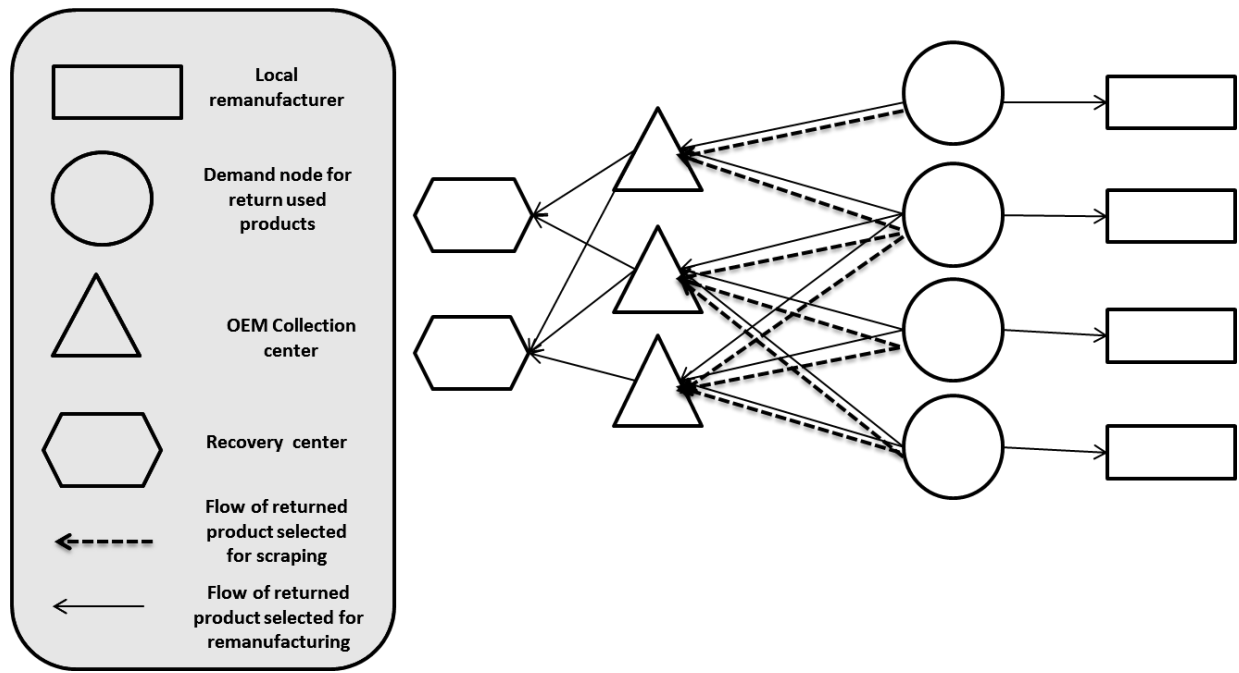

Figure 1. A network of facilities consists of demand nodes, collection centers, recovery centers and a local remanufacturer.

The network, shown in Figure 1, is a single product network consisting of local remanufacturers, demand nodes for the return of used products, OEM collection centers, and recovery centers. As illustrated in Figure 1, returned products are potentially available from the demand nodes. The local remanufacturer tries to collect these products by offering competitive prices and providing nearby collection centers. The OEM should provide a competitive service for the collection of returned products, taking the local remanufacturers into consideration.

The returned products are classified into two groups, based on their quality and designated for either scrapping or recovery. Dashed arrows show the flow of products selected for scrapping and continuous arrows show the flow of returned products for remanufacturing. Scrapped products are sold in OEM collection centers and the products chosen for recovery are shipped to recovery centers.

We have used a cumulative distribution function, because our aim is to find a cut-off point at which to separate the qualities of products into those that are suitable for remanufacturing and those that should be scrapped. As in Galbreth and Blackburn (2006), it is assumed that the cumulative distribution function for the quality of returned products is given. The cumulative distribution function will be used to determine the percentage of products that should be remanufactured or scrapped. Since we have incorporated the quality level into the remanufacturing cost, we should multiply a fixed value for the remanufacturing cost ($C_{rem}$) by the coefficient of the remanufacturing cost ($h_q$) in order to calculate the final cost of remanufacturing.

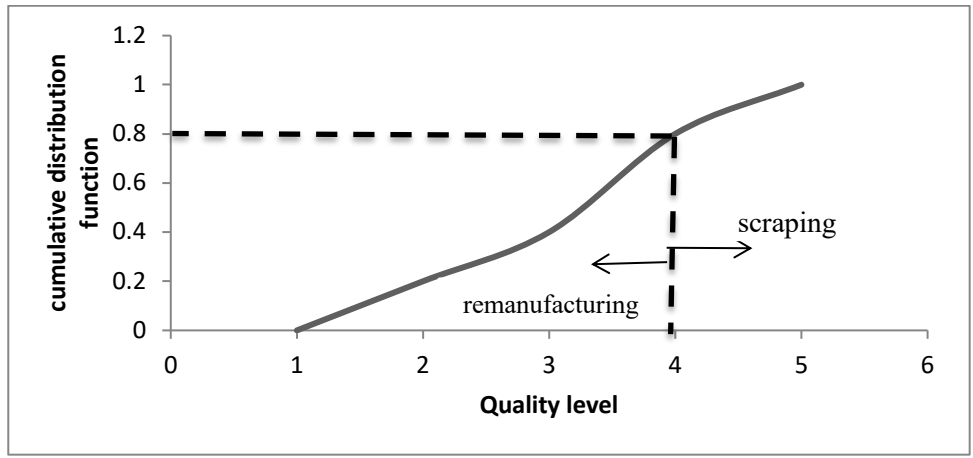

Figure 2. Cumulative distribution functions of returned products quality.

Customers choose collection centers, based on the offered prices and their distance. Since it is assumed that local remanufacturers already exist in the market, they will try to affect the customers' utility function by offering competitive prices and providing convenient locations for collection. The utility parameter $u_k$, which is determined by local remanufacturers, is included in the model. The fraction of customers at demand node $k$ that choose an original manufacturer collection center (set $E$) after considering the local remanufacturer's offered prices and collection centers (set $E^{'}$) is calculated using equation (1) (McFadden, 1973):

$$x_{ki} = \left[\frac{e^{v_k - d_{ki} y_{ki}}}{\sum_{r \in E \cup E'} e^{v_k - d_{ki} y_{ki}}}\right] \qquad \forall i \in N, i \in E \cup E', \qquad (1)$$

where $x_{ki}$ is the percentage of customers at demand node $k$ who are willing to go to the OEM collection center located at $i$. The complete list of variables and notation are provided in the later page. Since it is assumed that the local remanufacturer is already present in the market, we define $u_k$ as follows, in order to simplify equation (1),

$$u_k = \sum_{r \in E \cup E'} e^{v_k - d_{ki} y_{ki}}. \qquad (2)$$

We substitute this value in for the local remanufacturer's impact at demand node $k$. Thus, the fraction of costumers who choose the OEM's collection centers is given by:

$$x_{ki} = \left[\frac{e^{v_k - d_{ki} y_{ki}}}{e^{v_k - d_{ki} y_{ki}} + u_k}\right] \qquad (3)$$

In today's unstable world, businesses and, more specifically, supply chains are becoming increasingly global, and the industrial environment is heavily affected by uncertainty, which can potentially turn into unexpected disruptions (McCormack et al., 2008). Basically quantities that are not measurable are not manageable. The purpose of risk modeling is to gain insight into the system performance, represent or express the uncertainties, identify the risk contributors and see the effect of changes.

CVaR is an extension of VaR which has been widely used in economical situations (Sarykalin et al., 2008). To gain a better conception of CVaR, we begin by discussing VaR. Let $\alpha$ be the confidence level of predetermined probability. Then $\alpha$-VaR is the percentile of the loss distribution, or in other words, "$\alpha$-VaR is the smallest value for which the probability of loss exceeds or equals to this value is not larger than $1-\alpha$" (Babazadeh et al., 2015).

The measure, $\alpha$-CVaR is the minimization of "the expected value of the costs in the $(1-\alpha)$ 100% worst cases" (Schultz and Tiedemann, 2006). CVaR measures the risk of an investment in a conservative way by considering the maximum lost (Babazadeh et al., 2015).

VaR and CVaR are both suitable metrics for risk measurement. We used CVaR, because CVaR includes more information than VaR (determines the quantity of excess loss) and is also linear, convex and continuous, so it is much easier to compute (Soleimani et al., 2014). Following the work of (Rockafellar and Uryasev, 2002), the scenario-based calculation of CVaR is proposed.

It is assumed that positive values of $f(x,\omega)$ represent losses. In addition, we assume that $y$ has a finite discrete distribution with $N$ realizations and corresponding probabilities given by $\pi_\theta$ for $\omega_\theta$, $\theta = 1, .., N$, ($\theta$ is representative of a particular scenario) with $\pi_\theta > 0$, for all $\theta$, where $\sum_\theta \pi_\theta = 1$.

For $f(x,\omega)$, the $\alpha$-CVaR can be calculated using the following minimization formula:

$$F_\alpha(x, \omega) = \eta + \frac{1}{1-\alpha} E(\max\{(f(x,\omega) - \eta), 0\}). \qquad (4)$$

Let $\alpha$-CVaR for the loss random variable $f(x,\omega)$ be denoted by $\psi_\alpha(x)$. Then the $\alpha$-CVaR equation (equation (4)) can be restated as:

$$\psi_\alpha(x) = \min \{\eta + \frac{1}{1-\alpha} E(\max\{(f(x,\omega) - \eta), 0\})\} \quad (5)$$

Additional variables $Z_\theta$ representing $(\max\{(f(x,\omega) - \eta), 0\})$ for all $\theta = 1,..,N$ are introduced and a well-known idea in linear programming is incorporated to transform this non-linear programming problem into a linear programming problem. By expanding the expected value of $(\max\{(f(x,\omega) - \eta), 0\})$ for all scenarios, the following equivalent linear programming problem is obtained (Kall and Mayer, 2005):

$$\psi_\alpha(x) = \gamma + \frac{1}{(1-\alpha)} \sum_{\theta=1}^{N} \pi_\theta Z_\theta \quad (6)$$

$$f(x,\omega) - \eta - Z_\theta \leq 0 \quad \forall \theta \quad (7)$$

$$Z_\theta \geq 0 \quad \forall \theta \quad (8)$$

The following assumptions are used in our model:

i. The potential locations of recovery, collection and customers' demand nodes are known.
ii. There is no limitation on the capacity of material flowing through the network.
iii. There is no limitation on the capacity of the collection centers.
iv. The cost values (i.e. fixed, collection, recovery, and shipping costs) are known.
v. Any returned product, which is chosen to scrap, has a specified value.
vi. The cost of remanufacturing differs with the quality of the returned products.
vii. Scrapped returned products are sold in collection centers.
viii. Any demand node is assigned to just one OEM collection center to return products.
ix. The cumulative distribution function for the quality of returned products is known.

The following notation is used in formulation of the model

**Sets:**

i      index of potential locations for collection centers

j      index of recovery centers

k      index of demand nodes for returning products

$\theta$      index of scenarios

q      index of quality

**Parameters:**

$\lambda$      Trade-off coefficient representing the exchange rate of mean cost for risk

$\alpha$      Confidence level

$\pi_\theta$      Probability of scenario $\theta$ occurring

$R_{k\theta}$      Quantity of returned products in scenario $\theta$

P      The value of the remanufactured product

C:      The value of the scrapped product

$h_q$:      The coefficient of the remanufacturing cost for quality $q$

$C_{rem}$      Fixed cost of remanufacturing

$u_k$      The utility of node $k$

$CC_j$      Maximum capacity of recovery center $j$

$d_{ki}$      Cost of traveling from demand node $k$ to collection center $i$

$a_{ij}$      The cost of shipping returned products from collection center $i$ to recovery center $j$

$f_i$      Fixed cost for establishing collection center $i$

$d_{ki}$      Cost of travelling between demand node $k$ and collection center i

**Decision variables**

$v_k^r$      The financial incentive offered to customers in demand node $k$ for returned products suitable for remanufacturing.

$v_k^s$      The financial incentive offered to customers in demand node $k$ for returned products suitable for scrapping.

$D_{kiq\theta}^s$ The quantity of returned products in scenario $\theta$ attracted from demand node $k$ by collection center $i$ for remanufacturing when quality level $q$ is selected as the cut-off

$D_{kiq\theta}^r$ The quantity of returned products in scenario $\theta$ attracted from demand node $k$ by collection center $i$ for scrapping when quality level $q$ is selected as the cut-off

$\beta_q$     Percentage of returned products selected for remanufacturing

$U_{ijq\theta}$     The quantity of returned products in scenario $\theta$ shipped from collection center $i$ to recovery center j for remanufacturing when quality level $q$ is selected as cut-off

$$q_l = \begin{cases} 1 \text{ if quality level l selected as cut\_off point} \\ 0 \text{ otherwise} \end{cases}$$

$$y_i = \begin{cases} 1 \text{ if collection center opened at location } i \\ 0 \text{ otherwise} \end{cases}$$

$$y_{ki} = \begin{cases} 1 \text{ if the path between demand node } k \text{ and collection center } i \text{ is activated} \\ 0 \text{ otherwise} \end{cases}$$

$$y_{ij} = \begin{cases} 1 \text{ if the path between collection center } i \text{ and recovery center } j \text{ is activated} \\ 0 \text{ otherwise} \end{cases}$$

*Maximize:*

$$\sum_{q \in Q} \sum_{sc \in SC} \sum_{k \in K} \sum_{i \in I} P \, D^r_{kiq\theta} \pi_\theta + \sum_{q \in Q} \sum_{sc \in SC} \sum_{k \in K} \sum_{i \in I} C \, D^s_{kiq\theta} \pi_\theta$$
$$- (1 + \lambda) \sum_{i \in I} f_i y_i$$
$$- \left( \sum_{q \in Q} \sum_{sc \in SC} \sum_{k \in K} \sum_{i \in I} v^r_k D^r_{kiq\theta} \pi_\theta \right.$$
$$+ \sum_{q \in Q} \sum_{sc \in SC} \sum_{k \in K} \sum_{i \in I} v^s_k D^s_{kiq\theta} \pi_\theta \quad (9)$$
$$+ \sum_{q \in Q} \sum_{sc \in SC} \sum_{i \in I} \sum_{j \in J} U_{ijq\theta} \, a_{ij} y_{ij} \pi_\theta$$
$$\left. + \sum_{q \in Q} \sum_{sc \in SC} \sum_{i \in I} \sum_{j \in J} U_{ijq\theta} \, h_q C_{rem} \, \pi_\theta \right)$$
$$- \lambda \left( \gamma + \frac{1}{1 - \alpha} \sum_{\theta \in \Theta} \pi_\theta z_{sc} \right)$$

*Subject to:*

$$D^r_{kiq\theta} = R_{k\theta} q_l \beta_q \left[ \frac{e^{v^r_k - d_{ki} y_{ki}}}{e^{v^r_k - d_{ki} y_{ki}} + u_k} \right] \quad \forall \, k \in K, i \in I, \theta \in \Theta \quad (10)$$

$$D^s_{kiq\theta} = R_{k\theta} q_l (1 - \beta_q) \left[ \frac{e^{v^s_k - d_{ki} y_{ki}}}{e^{v^s_k - d_{ki} y_{ki}} + u_k} \right] \quad \forall \, k \in K, i \in I, \theta \in \Theta \quad (11)$$

$$v^r_k > d_{ki} y_{ki} \quad \forall \, k \in K, i \in I \quad (12)$$

$$v^s_k > d_{ki} y_{ki} \quad \forall \, k \in K, i \in I \quad (13)$$

$$\sum_{j} U^r_{ijq\theta} y_{ij} = \sum_{k \in K} D^r_{kiq\theta} \qquad \forall\, i \in I, \theta \in \theta \tag{14}$$

$$D^r_{kiq\theta}(y_{ki} - 1) \geq 0 \qquad \forall\, k \in K, i \in I, \theta \in \theta \tag{15}$$

$$D^s_{kiq\theta}(y_{ki} - 1) \geq 0 \qquad \forall\, k \in K, i \in I, \theta \in \theta \tag{16}$$

$$\sum_{i \in I} U_{ij\theta} y_{ij} \leq cc_j \qquad \forall\, j \in J, \theta \in \theta \tag{17}$$

$$y_{ki} \geq y_i p_{ki} \qquad \forall\, k \in K, i \in I \tag{18}$$

$$y_{ij} \geq y_i p_{ij} \qquad \forall\, i \in I, j \in J \tag{19}$$

$$Z_\theta \geq 0 \qquad \forall\, \theta \in \theta \tag{20}$$

$$Z_{sc} \geq \left( \sum_{\theta \in \theta}\sum_{k \in K}\sum_{i \in I} v^r_k D^r_{kiq\theta} \pi_\theta + \sum_{\theta \in \theta}\sum_{k \in K}\sum_{i \in I} v^s_k D^s_{kiq\theta} \pi_\theta \right.$$
$$+ \sum_{\theta \in \theta}\sum_{i \in I}\sum_{j \in J}(U^r_{ij\theta}) a_{ij} y_{ij} \pi_\theta \qquad \forall\, \theta \in \theta \tag{21}$$
$$\left. + \sum_{q \in Q}\sum_{\theta \in \theta}\sum_{i \in I}\sum_{j \in J} U^r_{ijq\theta} h_q C_{rem} \pi_\theta \right) - \gamma$$

$$y_{ij}, y_{ki}, y_i \in \{0,1\} \qquad \forall\, k \in K, i \in I, j \in J \tag{22}$$

$$v^d_k, v^r_k \geq 0 \qquad \forall\, k \in K \tag{23}$$

Objective function 7 consists of seven parts. In the first and second parts, revenue is estimated from remanufacturing and scrapping returned products. The third part shows the transportation cost of returned products from collection centers to recovery centers. Financial incentive costs are considered in the fourth and fifth parts.

Constraint 10 shows the demand for returning products collected for remanufacturing in scenario $\theta$. Constraint 11 shows the demand for returning products collected for scrapping in this scenario. Constraints 12 and 13 are financial incentives. Constraint 14 is the balance equation for remanufacturing returned products in collection centers. Constraints 15 and 16 ensure that products flow, only if the path between demand nodes and collection centers exists. Constraint 17 is a capacity constraint for recovery centers. Constraints 18 and 19 are the path constraints. Constraints 20 and 21 are risk parameters which were already defined in equations 7 and 8. Finally, Constraints 22 and 23 are defining logical constraints for the decision variables.

## 4. Numerical Example

In this section, we provide a numerical example to demonstrate the capability of our model. Consider an OEM that has decided to collect returned products. The market consists of six demand nodes for returning products, each with a specified demand. The cost of remanufacturing ($C_{rem}$) is 20 units, the value of the scrapped products (C) is 5 units, and the maximum capacity of recovery centers is 2000. Table 2 shows the complete data for this example.

Table 2. Values of the Numerical Example Parameters

| Returning location k | $d_{ki}$ collection center i | | | $u_k$ | Quality level q | $\beta_q$ | $h_q$ | Collection center | $f_i$ | Scenario $\theta$ | $Pr_\theta$ | $R_{k\theta}$ |
|---|---|---|---|---|---|---|---|---|---|---|---|---|
| | 1 | 2 | 3 | | | | | | | | | |
| 1 | 1.7 | - | 1.9 | 40 | 1 | 0 | 0.05 | 1 | 1000 | 1 | 0.25 | Uniform (0,500) |
| 2 | 1.5 | 1.13 | - | 60 | 2 | 0.2 | 0.1 | 2 | 1200 | 2 | 0.5 | Uniform (500,1000) |
| 3 | - | 1.9 | - | 60 | 3 | 0.4 | 0.15 | 3 | 1100 | 3 | 0.25 | Uniform (1000,1500) |
| 4 | 1.15 | - | 1.1 | 40 | 4 | 0.8 | 0.4 | | | | | |
| 5 | - | 1.9 | 1.12 | 80 | 5 | 1 | 0.5 | | | | | |
| 6 | 1.12 | 1.13 | - | 40 | | | | | | | | |

The model was coded using the Generalized Algebraic Modeling System (GAMS). The proposed solution algorithms were run on a computer with 2.54 GHz processor and 8 GB RAM.

In this example, we considered the uniform distribution function for the quality of the returned products. Three scenarios were considered to represent the quantity of returned products.

We analyzed the impact of $\alpha$ and $\lambda$ to show the effect of uncertainty on the model. Three cases for $\alpha$ (0.99, 0.95 and 0.9) and six for $\lambda$ (0, 0.3, 0.6, 1, 3, and 10) were considered. Tables 3, 4, 5, 6, and 7 show the results obtained for different values of $\alpha$ and $\lambda$. The results for the case with no risk are shown in table 8. Different values of risk parameter are used to show the flexibility and usefulness of the presented model.

The other main factor that we have considered is the quality of products and variations of them in the model. We have considered three different qualities for the returned products under different conditions. Three different levels of risk are considered for each $\lambda$ as shown in tables 3 to 7. In total, 45 different scenraios are analyzed.

Table 3. Values of Decision Variables for λ =0.3.

| | | λ =0.3 | | | | | |
|---|---|---|---|---|---|---|---|
| | | α=0.99 | | α=0.95 | | α=0.9 | |
| | | $v_k^r$ | $v_k^s$ | $v_k^r$ | $v_k^s$ | $v_k^r$ | $v_k^s$ |
| Collection center | 1 | 3.274 | 1.72 | 4.012 | 1.7 | 3.216 | 1.9 |
| | 2 | 3.268 | 1.68 | 4.078 | 1.5 | 4.038 | 1.5 |
| | 3 | – | – | – | – | 3.518 | 1.247 |
| | 4 | 3.147 | 1.15 | 3.805 | 1.07 | 2.391 | 1.616 |
| | 5 | – | – | – | – | 3.876 | 1.741 |
| | 6 | 3.139 | 1.12 | 3.793 | 1.06 | 3.265 | 2.872 |
| $q_l$ | | q3 | | q2 | | q3 | |
| $y_i$ | | $y_1$ | | $y_1$ | | $y_1, y_2$ | |
| Objective Values | | 3265.8 | | 4910.44 | | 5209.56 | |

Table 4. Values of Decision Variables for λ =0.6

| | | λ =0.6 | | | | | |
|---|---|---|---|---|---|---|---|
| | | α=0.99 | | α=0.95 | | α=0.9 | |
| | | $v_k^r$ | $v_k^s$ | $v_k^r$ | $v_k^s$ | $v_k^r$ | $v_k^s$ |
| Collection center | 1 | 4.012 | 1.731 | - | - | 3.011 | 1.7 |
| | 2 | 4.078 | 1.64 | 3.52 | 1.13 | 3.041 | 1.5 |
| | 3 | - | - | 3.152 | 1.147 | - | - |
| | 4 | 3.805 | 1.26 | - | - | 2.907 | 1.15 |
| | 5 | - | - | 3.217 | 1.012 | - | - |
| | 6 | 3.793 | 1.24 | 3.271 | 1.314 | 2.9 | 1.12 |
| $q_l$ | | q2 | | q3 | | q3 | |
| $y_i$ | | $y_1$ | | $y_2$ | | $y_1$ | |
| Objective Values | | 3111.01 | | 4658.38 | | 4849.23 | |

Table 5. Values of Decision Variables for λ =1

|  |  | λ =1 | | | | | |
|---|---|---|---|---|---|---|---|
|  |  | α=0.99 | | α=0.95 | | α=0.9 | |
|  |  | $v_k^r$ | $v_k^s$ | $v_k^r$ | $v_k^s$ | $v_k^r$ | $v_k^s$ |
| Collection center | 1 | - | - | 3.011 | 1.7 | 4.481 | 1.241 |
| | 2 | 3.108 | 1.74 | 3.041 | 1.56 | 5.563 | 1.5 |
| | 3 | 3.224 | 1.9 | - | - | - | - |
| | 4 | - | - | 2.907 | 1.35 | 5.044 | 1.15 |
| | 5 | 3.252 | 1.9 | - | - | - | - |
| | 6 | 3.019 | 1.73 | 2.9 | 1.62 | 5.023 | 1.458 |
| $q_l$ | | q3 | | q3 | | q2 | |
| $y_i$ | | $y_2$ | | $y_1$ | | $y_1$ | |
| Objective Values | | 2689.35 | | 4129.32 | | 4452.35 | |

Table 6. Values of Decision Variables for λ =3

|  |  | λ =3 | | | | | |
|---|---|---|---|---|---|---|---|
|  |  | α=0.99 | | α=0.95 | | α=0.9 | |
|  |  | $v_k^r$ | $v_k^s$ | $v_k^r$ | $v_k^s$ | $v_k^r$ | $v_k^s$ |
| Collection center | 1 | 3.628 | 2.24 | 4.011 | 2.51 | 4.874 | 2.54 |
| | 2 | 3.451 | 2.85 | 3.857 | 1.97 | 5.652 | 2.64 |
| | 3 | - | - | - | - | - | - |
| | 4 | 3.512 | 2.12 | 3.726 | 1.34 | 5.124 | 2.61 |
| | 5 | - | - | - | - | - | - |
| | 6 | 3.248 | 1.89 | 3.241 | 2.15 | 5.241 | 2.87 |
| $q_l$ | | q2 | | q2 | | q2 | |
| $y_i$ | | $y_1$ | | $y_1$ | | $y_1$ | |
| Objective Values | | 2214.25 | | 3527.41 | | 34128.35 | |

Table 7. Values of Decision Variables for λ =10

| | | λ =10 | | | | | |
| --- | --- | --- | --- | --- | --- | --- | --- |
| | | α=0.99 | | α=0.95 | | α=0.9 | |
| | | $v_k^r$ | $v_k^s$ | $v_k^r$ | $v_k^s$ | $v_k^r$ | $v_k^s$ |
| Collection center | 1 | - | - | - | 3.85 | - | 3.21 |
| | 2 | - | - | - | 2.74 | - | 3.47 |
| | 3 | - | - | - | - | - | - |
| | 4 | - | - | - | 2.64 | - | 3.12 |
| | 5 | - | - | - | - | - | - |
| | 6 | - | - | - | 3.18 | - | 3.31 |
| $q_l$ | | – | | q1 | | q1 | |
| $y_i$ | | – | | $y_1$ | | $y_1$ | |
| Objective Values | | – | | 1274.34 | | 1385.54 | |

In table 3, risk-seeking behavior (decreasing the values of α and λ) leads to the establishment of two collection centers ($y_1$,$y_2$) by the model. So, all the demand nodes have been covered. Quality level three has been selected as the cut-off point. The relevant results and offered prices are given in table 3.

To obtain the results given in table 4, λ was increased to 0.6 while α was set equal to 0.9. In this case, the model established just collection center $y_1$ and offered lower prices than the former state. When λ was increased, the model behaved more conservatively and reduced the costs by establishing one collection center instead of two. When the λ value was increased to 1 in table 5 (α=0.9), the model selected level 2 of quality as the cut-off point while it established only one collection center, $y_1$. So it was preferable to collect products of better quality and offer higher prices. This confirms that increasing λ made the model risk-averse.

As discussed before, α represents the confidence level. There is no doubt that increasing the confidence level will make the model more risk averse. This is evident from table 3 in which, for α=0.95, the model first selected level 2 of quality as the cut-off point, while it established only collection center $y_1$. As can be seen from table 3, when the confidence level is raised even further (α=0.99), the model behaves more risk-aversely by selecting level 3 of quality and offering lower prices.

We cannot judge the risk-aversive behavior of the model just from the offered prices, but should also consider the quality, price and location of collection centers simultaneously. As we discussed earlier, increasing the confidence level should make the model more risk-averse while other factors such as quality and locations are fixed. Therefore, the offered prices should decrease. We can see from table 3 that, although α increased from 0.9 to 0.95, $v_k^r$ increased. This makes sense because, in this scenario, the model removed one of the collection centers and also selected quality level 2. This means that the model selected a better level of quality and should offer higher prices. Similar results were obtained in other cases, as can be seen from tables 4, 5, 6, and 7.

By increasing λ to 10, when α=0.95 and α=0.9, the model selected all of the products for scrapping and offered more prices comparable with the lower values of λ. This is the highest value which we have considered for scrapping. For α=0.99, the OEM does not enter the market and there is no quality cut-off point.

Table 8: Value of decision variables for λ =0

|  |  | $v_k^r$ | $v_k^s$ |
|---|---|---|---|
| Collection center | 1 | 3.012 | 1.254 |
|  | 2 | 3.658 | 1.784 |
|  | 3 | 3.487 | 1.247 |
|  | 4 | 2.487 | 1.482 |
|  | 5 | 3.613 | 1.269 |
|  | 6 | 2.987 | 1.526 |
| $q_l$ |  | q4 | |
| $y_i$ |  | $y_1, y_2$ | |
| Objective Values |  | 6248.24 | |

Table 8 provides the results without risk (λ =0). In this state, the model selected quality level 4 and two collection centers for collecting more returned products. The value of the objective function increased, since it tried to collect more returned products. According to the objective function, when λ =0, we need not consider the value of the confidence level, α.

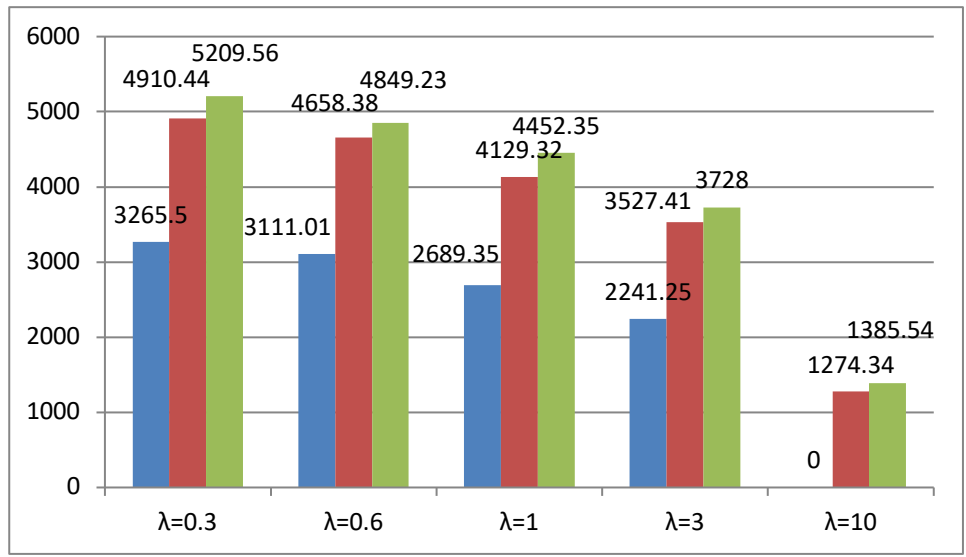

Fig3. Value of objective function for different α (■α =0.99, ■ α=0.95, ■α=0.9) and λ.

## 5. Model performance

In this section, in order to show the effect of using a risk measure in our model, the proposed mean-risk model is compared to the risk neutral model through VSS (the value of a stochastic solution which measures the average cost of ignoring uncertainty in risk neutral conditions).

In this paper three measures have been calculated:

i. MRRP (mean-risk recourse problem) = $\min\{E(f(x, \omega)) + \lambda CVAR_\alpha(f(x, \omega))\}$, $x \in R_n$
ii. MREV (mean risk expected value) = $E\{(f(\bar{x}(\bar{\varepsilon}), \omega)) + \lambda CVAR_\alpha(f(\bar{x}(\bar{\varepsilon}), \omega))\}$, where $\bar{x}(\bar{\varepsilon})$ is the optimal solution of the expected value problem, also $\bar{\varepsilon} = E(\bar{\varepsilon}(\omega))$
iii. MRVSS(Mean risk value of stochastic solution)= MRRP-MREV

MRVSS is an index for measuring the possible gain from solving the stochastic model. By increasing the level of risk-aversion (increasing in the values of λ α), the value of MRVSS is increased. This means that the quality of the solutions has increased and we should solve the risk-averse model instead of solving the expected value problem when we are confronted with risk parameters. In general, the value of MRRP and MREV increased when the level of risk-aversion decreased. As seen in table 9, when λ=0.3, and α decreased from 0.99 to 0.95, the value of MRRP increased from 3266 to 4910 and MREV increased from 124.3 to 351.2. MRRP and MREV touched 5210 and 602.5, respectively when α=0.9.

When α was fixed, increases in the value of λ led to decreases in MRRP and MREV. The 1st, 4th, 7th, 10th, and 13th instances show that while α=0.99, the value of MRRP decreased continuously from 3266 to 0. Also, MREV decreased from 124.3 when λ=0.3 to -5742 when λ=10.

The VSS measures the cost of ignoring uncertainty on average while making a decision. However, it only focuses on the risk-neutral approach.

For any mean-risk stochastic programming problem, we have MREV ≤ MRRP (Noyan, 2012).

Table 9. Values of MRVSS, MRRP, MREV

| alpha | 0.01 | 0.01 | 0.01 | 0.01 | 0.01 | 0.05 | 0.05 | 0.05 | 0.05 | 0.05 | 0.1 | 0.1 | 0.1 | 0.1 | 0.1 |
|---|---|---|---|---|---|---|---|---|---|---|---|---|---|---|---|
| lambda | 0.3 | 0.6 | 1 | 3 | 10 | 0.3 | 0.6 | 1 | 3 | 10 | 0.3 | 0.6 | 1 | 3 | 10 |
| MRRP | 3266 | 3111 | 2689 | 2214 | 0 | 4910 | 4658 | 4129 | 3527 | 1274 | 5210 | 4849 | 4452 | 3728 | 1386 |
| MREV | 124.3 | -246 | -1485 | -2018 | -4342 | 1351 | 814.5 | -57.7 | -1523 | -3818 | 1603 | 825.4 | -309 | -1225 | -3621 |
| MRVSS | 3142 | 3357 | 4175 | 4232 | 4342 | 3559 | 3844 | 4187 | 5051 | 5092 | 3607 | 4024 | 4762 | 4953 | 5007 |

As we see in figures 4, 5, and 6, changing the values of λ changed the size of MREV more than MRRP. The values of MREV and MRRP changed slowly for different values of α. The level of risk aversion was influenced simultaneously by the values of α and λ. MRVSS changed rapidly as λ increased from 0.6 to 3, but almost remained fixed after a specific value of λ was reached. The values of MRVSS depend on the decision-maker's behavior. Since MRVSS shows the difference between MRRP and MREV, and the decision maker was risk-averse, this difference was more noticeable. If the decision maker was more conservative, MRVSS changed slightly, because the risk affected the model at this stage. This affected MRVSS, so that fewer changes occurred for larger values of λ.

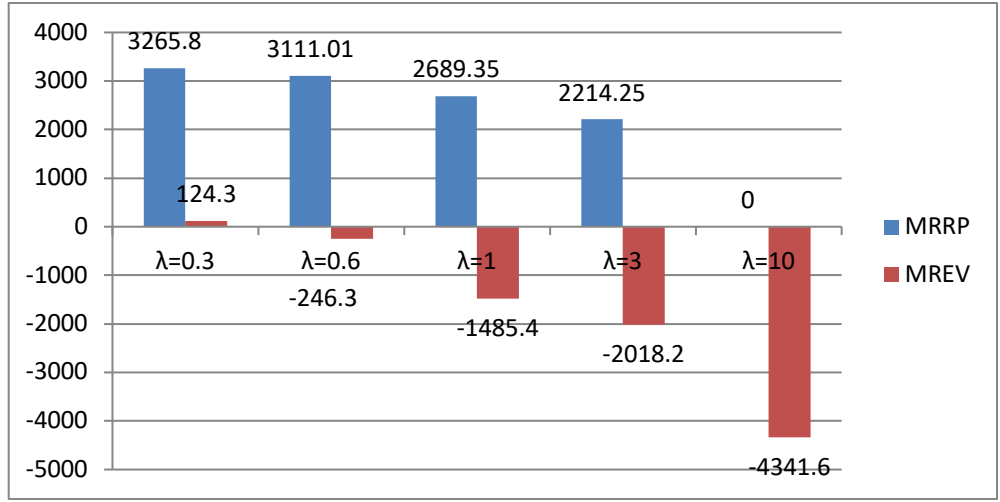

Fig4. Values of MRRP and MREV when α=0.99

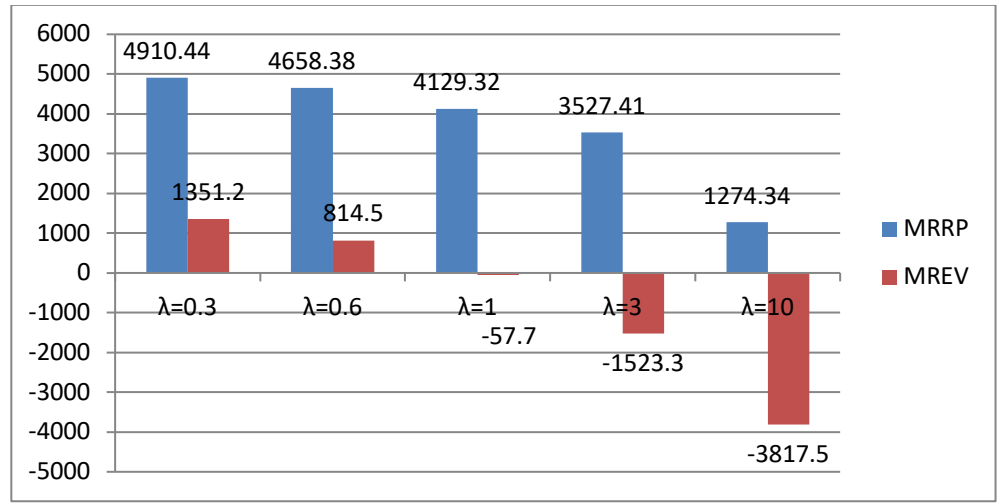

Fig5. Values of MRRP and MREV when α=0.95

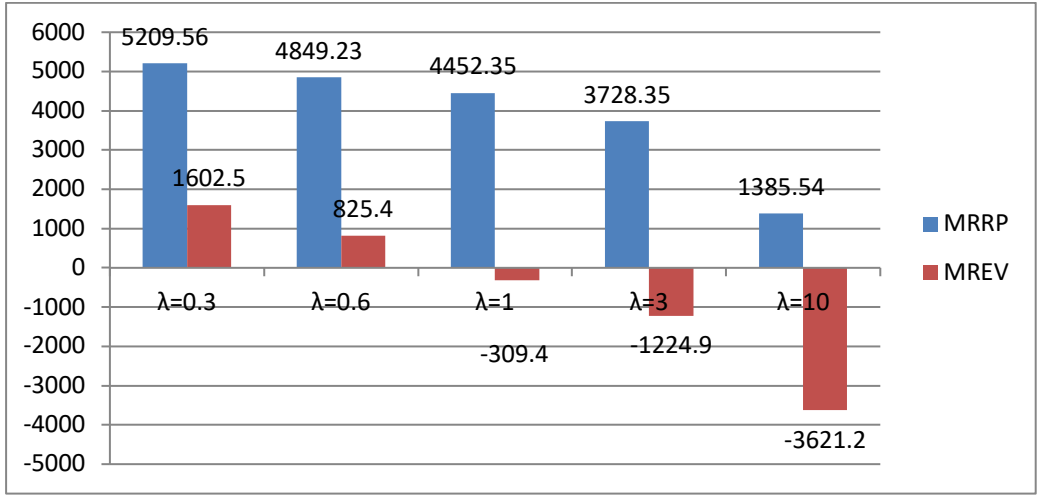

Fig6. Values of MRRP and MREV when α=0.9

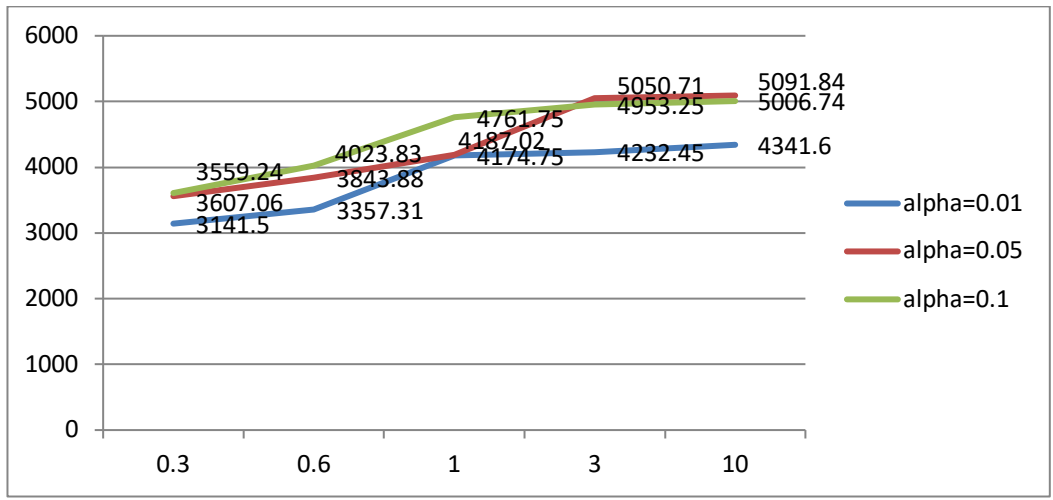

Fig7. Values of MRVSS for different values of α ( α =0.99, α=0.95, α=0.9), λ.

## 6. Conclusion

The model presented in this paper proposed a framework for designing reverse logistic networks. This model considered uncertainty in the quality and quantity of returned products. Decisions to remanufacture or scrap returned products were based on their quality. CVaR was applied to measure the effects of risk in the model and the performance of the model was analyzed for different values of α and λ (risk parameters).

The presented model is an MINLP model that was solved using GAMS. Our results imply that the model behaves more conservatively for increasing values of λ and α. In addition to the prices offered for returned products, the model can choose the location of collection centers and determine the level of quality for remanufacturing or scrapping in order to determine the best strategy in a volatile market. This model is useful and provides flexibility for managers, as it allows them to consider different values of α and λ, according to their circumstances, and to thereby make more reliable decisions.

This model can be used as the basis for many possible future research directions. Enhancements to consider inventory management costs, using game theory for pricing and considering uncertainty in other parameters such as transportation costs and capacity are some possible research directions that could be considered. Other exact solution methods could be employed to obtain the global optimum. Metaheuristics algorithms may also be useful for solving this model on a larger scale.